\documentclass[12pt]{article}
\usepackage{graphicx}
\usepackage[printonlyused]{acronym}
\usepackage{hyperref}
\usepackage{xcolor}
\usepackage[letterpaper, margin=1in]{geometry}
\usepackage{amsmath}
\usepackage{graphicx}
\usepackage{authblk}
 
\usepackage[font=itshape]{quoting}
\usepackage{float}
\usepackage{booktabs}
\usepackage{threeparttable}
\usepackage{tabularx}
\usepackage[symbol]{footmisc}
\let\OLDthebibliography\thebibliography
\renewcommand\thebibliography[1]{
  \OLDthebibliography{#1}
  \setlength{\parskip}{0pt}
  \setlength{\itemsep}{0pt plus 0.9ex}
}

\usepackage{booktabs}
\usepackage{threeparttable}
\usepackage{float}
\usepackage{enumitem}
\RequirePackage{amsthm,amsmath,amsfonts,amssymb}

\title{Priors from Envisioned Posterior Judgments: A Novel Elicitation Approach With Application to Bayesian Clinical Trials}
\author[1,2]{Yongdong Ouyang \footnote{Corresponding Author. Email: youyang2@buffalo.edu}}
\author[3]{Janice J Eng}
\author[4]{Denghuang Zhan}
\author[5,6]{Hubert Wong}
\author[ ]{The WnW Research Team \footnote{The WnW Research Team includes Sue Peters, Stanley H Hung, Mark T Bayley, Krista L Best, Louise A Connell, Sarah J Donkers, Sean P Dukelow, Victor E Ezeugwu, Marie-Hélène Milot, Brodie M Sakakibara, Lisa Sheehy, Hubert Wong, Yuwei Yang, Jennifer Yao, Janice J Eng}}
\affil[1]{Department of Biostatistics and Bioinformatics, Roswell Park Comprehensive Cancer Center, Buffalo, NY, USA}
\affil[2]{The State University of New York at Buffalo, Buffalo, NY, USA}
\affil[3]{Centre for Aging SMART and Department of Physical Therapy, University of British Columbia, Vancouver, BC, Canada}
\affil[4]{Global Statistical Sciences, Eli Lilly and Company, Little Island, Cork, Ireland}
\affil[5]{School of Population and Public Health, University of British Columbia, Vancouver, Canada}
\affil[6]{Centre for Advancing Health Outcomes, St Paul's Hospital, Vancouver, Canada}
\date{}

\begin{document}

\maketitle

\newpage
\begin{abstract}
\textbf{Background:} The uptake of formalized prior elicitation from experts in Bayesian clinical trials has been limited due to challenges such as complex statistical modeling, lack of practical tools, and the cognitive burden placed on experts arising from needing to quantify their uncertainty probabilistically. Existing methods also fail to address prior-posterior coherence, i.e., how do we ensure that the posterior distribution, obtained mathematically from combining the estimated prior with the trial data, reflects the expert's actual posterior beliefs?

\textbf{Method:} In this study, we propose a new elicitation approach that effectuates prior-posterior coherence and reduces cognitive burden. This is achieved by eliciting expert responses, comprising point estimates only, about envisioned posterior judgments under various data outcomes and inferring the prior distribution by minimizing discrepancies between these responses and expected responses derived from the posterior distribution. Via an iterative process, experts receive feedback on the degree of coherency of their responses, and are invited to revise their responses to achieve greater coherency.  The feasibility and potential value of this new approach are illustrated through an application to an ongoing trial.

\textbf{Results:} We involved 10 experts from Walk 'n watch trial research team. Experts were presented with 16 hypothetical outcome scenarios to experts and elicit the priors followed by the developed elicitation framework. Following two rounds of elicitation, experts' judgments showed substantial improvement in coherency, demonstrating the practical applicability of the proposed elicitation approach.

\textbf{Conclusion:} The proposed method provides a practical solution to the challenges of formalized prior elicitation in Bayesian clinical trials by addressing prior-posterior coherence and reducing cognitive demands on experts.\\

\textbf{Keywords:} Prior elicitation, Bayesian, Decision making, Trial design, Informative prior, Prior-posterior coherence

\newpage
\section{Background}

Bayesian clinical trials have become more popular recently \cite{berry_bayesian_2006,berry_bayesian_2010,berry2013bayesian}. A key feature of the Bayesian approach is the potential to incorporate domain knowledge, previous data, or expert opinions through prior distributions on model parameters into the design and analysis to increase trial efficiency and obtain more accurate estimates of the intervention effects \cite{Gunn-Sandell_Bedrick_Hutchins_Berg_Kaizer_Carlson_2024,leeBayesian2012}. However, the adoption of Bayesian methods has been slow despite growing interest \cite{clark_why_2023}. The lack of agreement and methodology on how to choose prior distributions for model parameters is arguably the biggest hurdle to Bayesian analysis \cite{Siddiqueprior2023}. Prior elicitation is a scientific process that transforms domain knowledge, previous data, or expert judgments into well-defined prior distributions. It offers a solution to the prior specification problem, especially when limited data is available \cite{ohagan_uncertain_2006,ohagan_expert_2019}. Applied to clinical trials, elicitation involves engaging medical experts to assist them in summarizing their judgments on how well treatments work, based on available data and clinical experience, and communicating those judgments in a way that the results can be combined with trial data \cite{ijerph18041833}. However, a major impediment has been the availability, user acceptability, and difficulty of implementation of approaches to eliciting priors from experts.

Formalized prior elicitation approaches have been available for more than two decades \cite{ohagan_uncertain_2006}. However, uptake in these methods has been limited in practice due to the need for complex statistical modeling, lack of practical tools and implementation frameworks, and a shortage of trained of elicitation facilitators \cite{mikkola_prior_2023}. Current elicitation approaches include (1) direct elicitation of the probability distribution for the parameter(s), (2) indirect elicitation based on judgments about the plausibility of observable data, and (3) modeling of past experts' decisions \cite{Gosling2018,MORRIS20141,ijerph18041833,ouyang_increasing_2021}. These methods typically require experts to quantify their uncertainty using probabilistic language. However, while humans typically are comfortable with providing point estimates (i.e., “best guesses”), they are less comfortable with providing probabilistic statements regarding their uncertainty. For example, one may be comfortable stating ``I think there is a 20\% chance that it will rain tomorrow.”, but be uneasy about making a statement such as ``I am 90\% certain that the chance of rain tomorrow is between 10\% and 30\%.”) \cite{Tversky1974}. The judgments needed for statements like the latter impose a high cognitive burden on experts \cite{falconer_methods_2022}, which may require them to undergo extensive training to ensure they are adept at quantifying uncertainty using probabilistic statements and to mitigate potential cognitive biases \cite{bojke_developing_2021}.

One previously unacknowledged challenge is that existing methods do not address {\em prior-posterior coherence}, i.e., does the posterior distribution, obtained mathematically from combining the elicited prior with the trial data, reflect the expert’s actual posterior beliefs? If one proceeds axiomatically, then on observing the trial data, the posterior calculated from these data and the elicited prior describes what the expert {\em ought} to believe. However,  it is possible that after seeing the trial data, the expert may decide that this ``mathematical'' posterior does not describe the expert's actual posterior beliefs. If one accepts the axioms and data as correct (and assuming the expert had no reason to modify their prior), then a meaningful discrepancy between these posteriors would suggest that the previously elicited prior did not, in fact, reflect the true prior beliefs of the expert. Indeed, one could argue that prior-posterior coherence is the ``litmus test'' for judging whether the elicited prior truly reflects the expert’s prior beliefs. Yet existing elicitation methods focus solely on obtaining a prior distribution which is never tested in this way.

In this work, we propose a novel elicitation method, referred to as the ``prior-from-posteriors (PfP)” approach, which seeks to ensure prior-posterior coherence and reduce the cognitive burden of the expert. This method distinguishes itself from existing approaches in two significant ways: (1) We obtain from experts their envisioned posterior judgments given hypothetical outcome data from the trial, and (2) We collect only point estimates, excluding any judgments regarding certainty (uncertainty). In Section 2, we present our general mathematical framework for inferring the prior distribution, with a focus on elicitation for a univariate, normally distributed mean parameter. In Section 3, we illustrate the application of our method in a case study - the Walk 'n Watch (WnW) trial, including a description of the implementation aspects. In Section 4, we conclude with a discussion of potential ways to improve the elicitation process and to extend the framework to more general problems.

\section{Methods}
%Instead of asking for experts' judgments about the “most likely value” and their uncertainties， we are going to take a different approach. Instead of asking experts to make assessments of their prior judgments, we will ask them for their “posterior” assessments of the best estimate only (no probability intervals) when given various hypothetical trial data. From these best estimates, we can infer (statistically) what their prior distribution had to have been to best match these posterior judgments.

Formally, let $\theta$ denote the parameter(s) of interest (e.g., the mean difference in outcomes between intervention and control arms), $f(\theta; a)$ be a family of prior distributions for $\theta$ indexed by a set of fixed parameter values $a$, $f(Y \mid \theta)$ be the sampling model/likelihood function, and $f(\theta \mid Y; a) \propto f(Y \mid \theta)f(\theta; a)$ be the posterior distribution for $\theta$ given $Y$. Our proposed elicitation method consists of two steps:
\begin{itemize}
    \item Let $\mathbf{Y}_k$, $k = 1, \ldots, K$ denote a set of potential data outcomes (scenarios).  For each scenario, we elicit from the expert a point estimate $\tilde{\theta_k}$ for the posterior mean 
    \[
    \mu_k(a) \equiv E(\theta \mid \mathbf{Y}_k; a).
    \]
    \item Let $H_k(a) \equiv H\left(\tilde{\theta_k}, \mu_k(a)\right)$, where $H(u, v)$ is a chosen discrepancy function; e.g. squared difference $H(u, v) = (u - v)^2$.  Set the inferred prior distribution to $f(\theta; \hat{a})$ where $\hat{a}$ is the value of $a$ that minimizes the discrepancies on average across all $K$ scenarios. For example, the root mean square discrepancy (RMSD): 
    \[
    RMSD(a) = \sqrt{\frac{1}{K}\sum_{k=1}^K H\left(\tilde{\theta_k}, \mu_k(a)\right)} = \sqrt{\frac{1}{K}\sum_{k=1}^K\left(\tilde{\theta_k}- \mu_k(a)\right)^{2}}.
    \]
\end{itemize}

The value of $H_k(\hat{a})$ measures the ``incoherency'' of the expert’s judgment in scenario $k$ under the inferred prior and can be used by the expert to reflect on and refine their elicitation, and RMSD$(\hat{a})$ reflects the expert's achieved overall level of (in)coherence (with zero indicating perfect coherence).
 
In the case study presented below, the objective was to obtain a prior for the mean parameter (univariate) $\theta$ from a normal data model, i.e., $Y \mid \theta \sim \mathcal{N} \left( \theta, 1/\tau \right)$, where $\tau$ is the precision parameter. For the $k$-th scenario, experts were asked to provide their response $\tilde{\theta_k}$ given data $\mathbf{Y}_k$, comprised of a sample of $n_k$ observations for $Y$. We assumed a conjugate normal prior $\mathcal{N} \left( \mu_0, 1/\tau_0 \right)$ for $\theta$, and our elicitation goal was to infer the value of $a \equiv (\mu_0, \tau_0)$.  Under these assumptions, the posterior mean is $\mu_k(a) = (\tau_0 \mu_0 + n_k \bar{y} / s^2)/(\tau_0 + n_k / s^2)$, where $\bar{y}$ is the sample mean and $s^2$ is the sample variance.  With a squared difference as the discrepancy function, the inferred value for ($\mu_0$, $\tau_0$) from minimizing the RMSD is
$$(\hat{\mu_0}, \hat{\tau_0}) = \arg\min_{\mu_0, \tau_0}  \sqrt{\frac{1}{K}\sum_{k=1}^K\left(\tilde{\theta_k}- (\tau_0 \mu_0 + n_k \bar{y} / s^2)/(\tau_0 + n_k / s^2)\right)^{2}}$$
This optimization is straightforward as it involves only two variables and a well-behaved objective function.

%Let $\mu_0$ be the mean of the prior distribution, $\tau_0$ be its precision. The prior distribution of parameter of interest $f(\theta; a=\left[\mu_0, \tau_0  \right])$ can be denoted as $\mathcal{N} \sim (\mu_0, 1/\tau_0)$. This prior distribution is what we want to elicit from experts. Let $\mu$ be the mean of the likelihood, and $\tau$ be its precision (known). The posterior distribution is also normal with parameters:

%\begin{align*}
%  \text{Posterior Mean:}\quad \mu_n &= \frac{\tau_0 \mu_0 + n \tau \bar{y}}{\tau_0 + n \tau} \\
%  \text{Posterior Precision:}\quad \tau_n &= \tau_0 + n \tau
%\end{align*}

%Here, $n$ is the number of observations, and $\bar{y}$ is the sample mean.

%For any given hypothetical outcome data $\bar{y}$ and the sample size ($n$), experts could elicit a point estimate of the posterior mean $\theta(Y)$. This procedure will be repeated $K$ times (over $K$ different scenarios and pairs of $\bar{y}$ and $n$) such that we have enough information to find $a=\left[\mu_0, \tau_0 \right]$ by minimizing the squared difference between the elicited and expected posterior means.

\subsection{Case Study}
\subsubsection{Trial background}
Many individuals who have had a stroke experience impaired mobility, and a key therapeutic goal is to regain their ability to walk independently. The Walk ’n Watch (WnW) study (author JE is the trial's Principal Investigator (PI)) is a stepped-wedge cluster randomized trial (SW-CRT) that aims to investigate the impact of introducing a structured, progressive exercise program called the WnW protocol on the walking ability in adult patients recovering from stroke during their hospital inpatient rehabilitation \cite{peters_implementation_2023, peters_safety_2025}. The primary outcome in the WnW trial is the 6-Minute Walk Test (6MWT), which measures the distance that a participant can walk in 6 minutes. The trial involved twelve hospital sites, all of whom initially delivered standard care (phase I). Every four months, three hospitals were chosen randomly from those still delivering standard care to switch to delivering the WnW protocol (Phase II). A 2-week training period for the WnW protocol was scheduled during the transition from Phase I to Phase II, during which enrollment was suspended.

\subsubsection{Motivation for the elicitation}
The proposed primary analysis of WnW trial utilizes a frequentist linear mixed-effects model to evaluate the benefit of the WnW protocol versus Usual Care (UC) \cite{peters_implementation_2023}. However, the estimated treatment effects from a stepped-wedge design is confounded by time due to the strong correlation between treatment allocation and calendar time \cite{ouyang_explaining_2020, wong_randomization-induced_2019,OUYANG2021106255}. Hence the analysis model for a SW-CRT requires adjustment for a potential time trend \cite{HUSSEY2007182}, which severely reduces the power of the analysis irrespective of whether a time trend in fact is present \cite{ouyangSWsample2022,ouyang_explaining_2020}. Zhan et al. \cite{zhan_improving_2021} showed that incorporating an informative prior distribution on the time trend using a Bayesian analysis can mitigate this problem and recover some of the power. Hence, the key parameter of interest for elicitation here was the following: \textit{Over the course of one year during the trial, and excluding impacts of potential changes over time in participant characteristics that are included in the analysis model (age, sex, baseline 6MWT), by how much does the average distance walked on the 6MWT by participants enrolled at the start of the year compared to those enrolled at the end of the year?} 
The prior judgments of an expert could be influenced by various factors such as data on historical patterns of change \cite{psioda_bayesian_2019} and the expert's perception/experiences during the course of the trial. In this trial, an important consideration was the expert's judgment regarding the impact of changes in (usual) care on outcomes over time due to the COVID-19 pandemic. 

\subsubsection{Expert recruitment} 

For the WnW trial, we engaged ten clinicians who were involved in delivering or advising on the WnW protocol as experts. Two were physicians and eight were physical therapists.  Four of the clinicians were directly involved with one of the 12 WnW sites, while the other six were affiliated with three other rehabilitation sites.  All were familiar with current stroke rehabilitation practices and standards of care.  Eight of the clinicians had advanced research training (Masters/PhD), while two did not. While all had some prior clinical trial experience, the majority would employ a statistician to undertake statistical analyses. The elicitation was conducted prior to the experts seeing any outcome data from the WnW trial.

\subsection{Elicitation process}

Our elicitation process included the following steps:
\begin{enumerate}[noitemsep, topsep=0pt]
    \item Circulate introductory elicitation materials/information to the experts.
    \item Host an open online session for the experts to answer their questions about the elicitation.
    \item Collect initial elicitation responses from the experts and analyze to infer their prior distributions.
    \item Provide feedback on where their responses show poor coherency, with suggestions on how to improve coherency.
    \item Collect revised elicitation responses and infer revised prior distributions.
\end{enumerate}

\subsubsection{Introductory materials}

In this study, the only preparatory materials supplied to the experts was a single document describing the purpose of the exercise, and the 16 scenarios (hypothetical outcome datasets) for which we were eliciting a judgment about the expected change in mean outcome over one year (specifically in 2022 versus in 2021). Authors YO and HW created the scenarios, and JE tailored the language to make the requests easy for experts to understand. A copy of the full questionnaire is provided in the Supplementary Materials S1.

Unlike how elicitation is typically conducted, we provided no preparatory instruction or practice to the experts on probability concepts or making probability judgments. This was done intentionally, as one of our goals was to identify where experts encountered difficulties/needed additional support.

Table 1 summarizes the 16 hypothetical outcomes for the observed mean change in 6MWT from 2021 to 2022 ($\bar{y}$) among a sample of $n$ individuals in the UC group. In each scenario, the observed mean change was based on different sample sizes ($n$). Experts were requested to provide their envisioned estimate of the mean change in the 6MWT outcome ($\tilde{\theta}$) during the WnW trial from 2021 to 2022 in the study population if they were to observe each of 16 hypothetical outcomes.

\begin{table}[H]
  \centering
  \caption{Summary of hypothetical trial outcomes that were presented to experts for elicitation of posterior judgments}
%  \resizebox{\textwidth}{!}{
  \begin{tabular}{ccc}
  \toprule
    \hline
        & \multicolumn{2}{c}{ \textbf{Hypothetical outcome data}} \\ \cline{2-3}
    \textbf{Scenario} & \textbf{Sample Size} & \textbf{Mean 6MWT Change} \\ 
                      & ($n$)                      & ($\bar{y}$, meters)  \\ \hline
    \midrule
    \textbf{1}  & No Data  & Not Applicable  \\ \hline
    \textbf{2}  & 10   & 0    \\ \hline
    \textbf{3}  & 10   & +10  \\ \hline
    \textbf{4}  & 10   & +30  \\ \hline
    \textbf{5}  & 10   & $-10$  \\ \hline
    \textbf{6}  & 10   & $-30$  \\ \hline
    \textbf{7}  & 30   & 0    \\ \hline
    \textbf{8}  & 30   & +10  \\ \hline
    \textbf{9}  & 30   & +30  \\ \hline
    \textbf{10} & 30   & $-10$  \\ \hline
    \textbf{11} & 30   & $-30$  \\ \hline
    \textbf{12} & 100  & 0    \\ \hline
    \textbf{13} & 100  & +10  \\ \hline
    \textbf{14} & 100  & +30  \\ \hline
    \textbf{15} & 100  & $-10$  \\ \hline
    \textbf{16} & 100  & $-30$  \\ \hline
    \bottomrule
  \end{tabular}
  \label{tab:6MWT_changes}
\end{table}

\subsubsection{Preliminary Q \& A session}

Following the distribution of the preliminary materials, the trial PI hosted an optional online session to answer questions from the experts. The purpose was to allow experts the opportunity to clarify any ambiguities about what they were being asked to do or how to proceed.

\subsubsection{Initial elicitation data collection and analysis}
Experts were asked to complete the questionnaire by providing their best estimate for the mean change in 6WMT from 2021 to 2022 in the UC group (i.e., $\tilde{\theta_k}$) if they were to observe the outcomes from each of the 16 hypothetical scenarios. Our statisticians did not directly communicate with the experts during this stage; instead, the PI handled all communication. After receiving the elicited responses, we inferred the prior distribution for each expert and assessed their level of coherency using the method described in section 2. 

\subsubsection{Feedback to experts}

We reviewed the inferred priors and the level of coherency achieved by each expert and sought to identify contributors to poor coherency.  Based on this review, we provided written feedback to experts. This feedback comprised (1) general feedback given to all experts plus (2) individualized feedback specific to the responses provided by the expert. The general feedback included two generic examples explaining what ranges of responses would reflect coherency across scenarios and the underlying rationale.  This was followed by a suggestion for how to ensure coherency in general. (See Figure 1 for exact wording -- Note that the term `consistency' rather than `coherency' was used in communications with the experts as the former term was felt to be easier for them to understand.) 

\begin{figure}

\noindent\fbox{
\parbox[c]{\textwidth}{

We re-iterate that there is no right or wrong response to any given scenario. However, it is important that your judgments be reasonably consistent across different scenarios.  Here are two examples of what we mean by being consistent.\\

\textbf{Example 1:} Suppose that in scenario 1 (where you have not yet seen any outcome data), you think the average change on the 6MWT was $-10$m. In scenario 2, you are shown data that suggests no change. Then, a consistent response from you would be a value between $-10$m and 0m. It would be inconsistent to choose a value less than $-10$m since data showing no change would not support a decrease even larger than what you initially believed. Similarly, it would be inconsistent to choose a value greater than 0m since neither your initial judgment nor the data support a positive change.\\

\textbf{Example 2:} Suppose your responses were 0m (no change) for scenario 1 and $+5$m for scenario 3 (data show +10m based on 10 participants per arm). Then a consistent response for scenario 8 (data show $+10$m based on 30 participants per arm) would be a value in the range $+5$m to $+10$m. The reason is that even with only 10 participants per arm, there was sufficient evidence to shift the expert's judgment from 0m to 5m, so with stronger evidence (i.e., based on more participants), the shift should be even greater (up to the $+10$m shown in the data).\\

In summary, a simple approach to improving consistency is to ensure your response falls between what you judge the value to be when you have no data and the value shown in the data, with more weight being given to the data as the sample size increases.\\

}
}
\caption{General feedback provided to experts after the initial elicitation.}
\end{figure}

This general feedback was followed by individualized feedback, which included three items:
\begin{itemize}[noitemsep, topsep=0pt]
    \item A summary table showing the elicited responses from each scenario, the corresponding inferred ``best-fit" value, and the discrepancy between these two values     
    \item A plot of the best-fit values against the elicited values for the 16 scenarios
    \item A detailed narrative summary identifying the scenarios in which the responses were exhibited poor coherence, with an explanation of why.
\end{itemize} 

The written feedback documents were accompanied by an invitation for the experts to revise their responses to obtain greater coherence, as well as an offer from JE to discuss the feedback and to answer any questions prior to experts revising their responses. 

\subsubsection{Revising the elicitation}

In the initial elicitation, the scenarios were presented as “stand-alone” questions, i.e., without asking the experts to compare their responses across scenarios. The rationale was that experts could process stand-alone scenarios more easily. However, the results from the initial elicitation suggested this could have led to poor coherence across scenarios. Hence, for the revised elicitation, the experts were shown their responses for all scenarios together in table form, and were instructed to consider coherency across scenarios when revising their responses. After receiving the revised responses, the models were re-fitted to obtain the revised prior distributions.

%they commented that this enabled them to see more easily when their responses showed poor coherence and to achieve better coherence.After receiving the revised elicited values, we reproduce the individual feedback packages and send them back to experts.

\section{Results}
\subsection{Initial elicitation}

The inferred prior mean and standard deviation (SD) in metres for each expert are displayed in Table 2. This table was ordered from least coherent to most coherent, measured by RMSD. There was a large variation in both the means ($-30.8$m to 3.8m) and the SD (0.2m to 118.6m), indicating that beliefs about the time trend were highly diverse. However, the coherency measures indicate that there was poor coherence in the responses for many of the experts, which raises doubt as to whether these priors truly reflect the experts' beliefs. Figure 1 shows the observed versus predicted responses for two experts, one who achieved very good coherence and one who achieved poor coherence.

\begin{figure}[H]
    \centering
    \includegraphics[width=0.45\linewidth]{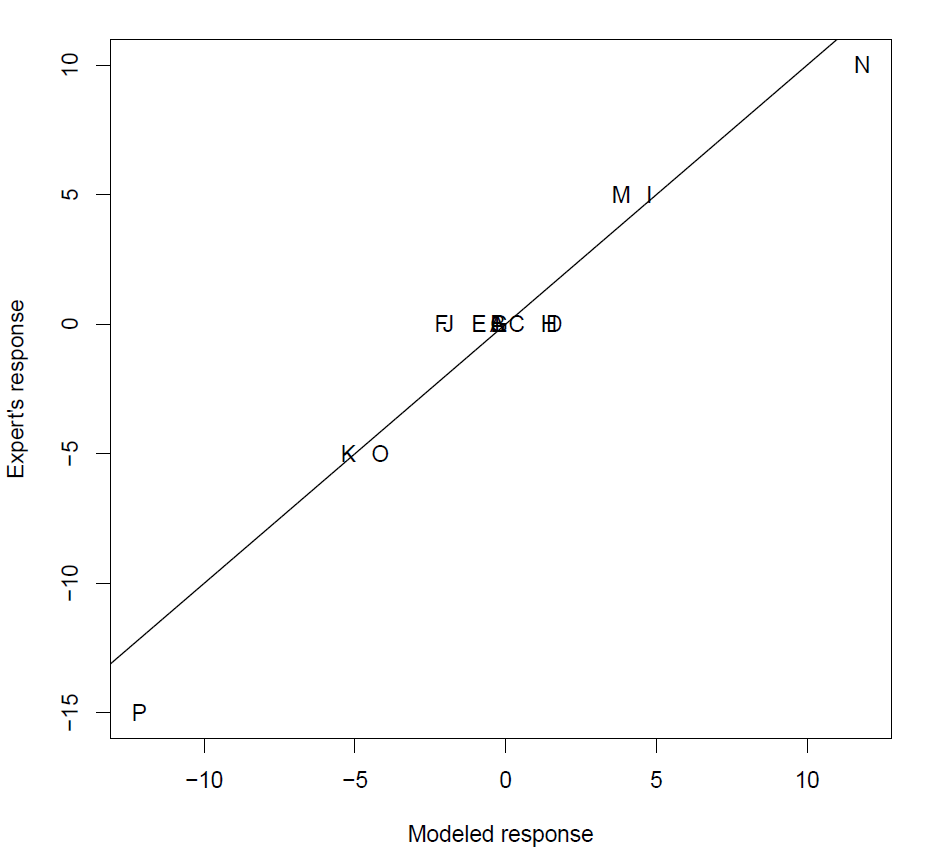}
    \includegraphics[width=0.45\linewidth]{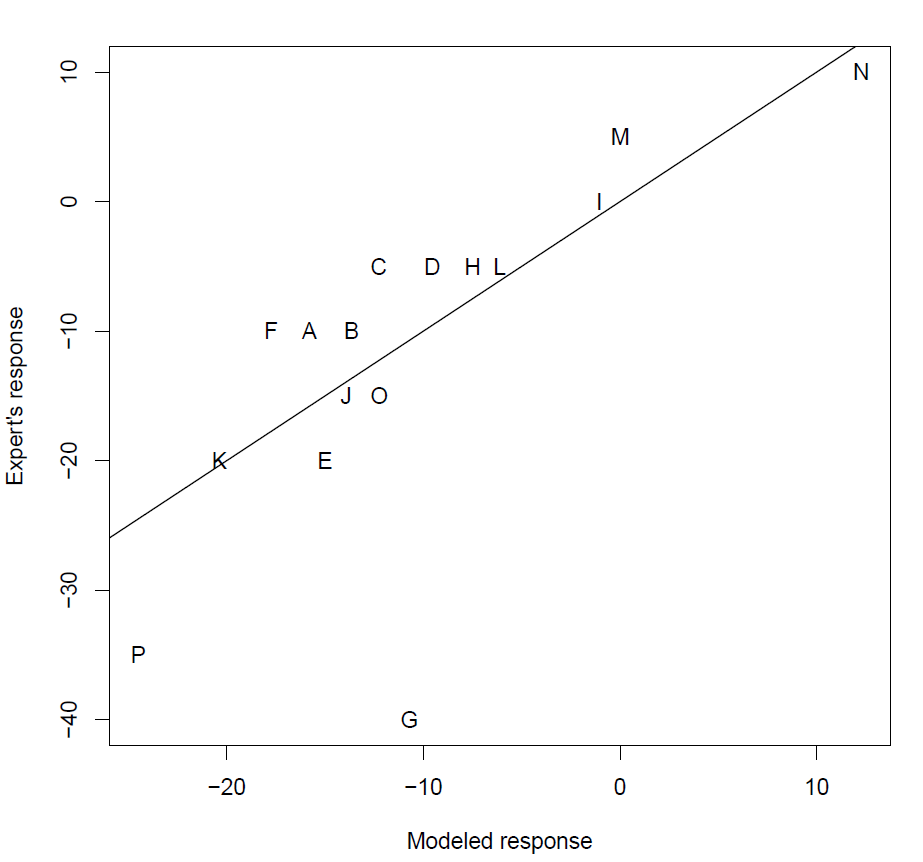}
    \caption{Comparison of elicited vs. best-fitted responses for experts 2 (left:2A) and expert 9 (right:2B). The labels A to N indicate the responses from scenarios 1 to 16, respectively, and the straight line is the line of equality between the elicited and best-fit response.}
    \label{fig:combined-label}
\end{figure}

Three experts (1, 2, 3) achieved a RMSD less than 3m. Note, though, that the expert who achieved the highest coherency (1) had an inferred prior mean of $-0.3$m and SD of 0.2m, indicating they were nearly certain that the mean change was very close to zero. For this expert, a concern would be whether this degree of certainty could be justified, despite the judgments being coherent. Examination of the individual responses for expert 10 strongly suggested the expert misunderstood what was requested, which would explain their large values for the prior SD and for the incoherency measure.

\begin{table}[H]
  \centering
  \caption{The mean, SD, and coherency (RMSD) of the inferred normal prior distribution for each expert (in meters).}
  \begin{tabular}{lrrr}
    \toprule
    \hline
    & \multicolumn{2}{c}{\textbf{Inferred Prior}} &  \\ 
       \midrule
    & \textbf{Mean} & \textbf{SD} & \textbf{RMSD} \\ \hline
    \textbf{Expert 1}  & $-0.3$  & 0.2   & 1.2  \\ \hline
    \textbf{Expert 2}  & $-0.3$  & 10.4  & 1.3  \\ \hline
    \textbf{Expert 3}  & $-30.8$ & 4.3   & 1.7  \\ \hline
    \textbf{Expert 4}  & $-9.4$  & 11.6  & 3.4  \\ \hline
    \textbf{Expert 5} & 3.8   & 20.9  & 5.6  \\ \hline
    \textbf{Expert 6}  & 0.0     & 32.2  & 6.0  \\ \hline
    \textbf{Expert 7}  & 0.0     & 15.2  & 6.1  \\ \hline
    \textbf{Expert 8}  & $-2.6$  & 30.6  & 6.6  \\ \hline
    \textbf{Expert 9}  & $-15.8$ & 16.0    & 8.7  \\ \hline
    \textbf{Expert 10}  & 0.9   & 118.6 & 21.7 \\ \hline
      \bottomrule
  \end{tabular}
\end{table}

An example of individual feedback packages can be found in the Supplementary Material S2. JE conducted virtual meetings in which three people attended to clarify the process and provide more detailed explanations.

\subsection{Revised elicitation}
Seven experts submitted revised responses. Table 3 presents each expert's revised inferred prior mean and SD, as well as the revised coherency measures. This table also showed the comparison of elicited distribution for each expert at the initial and revised stages. Except for the expert who had mis-understood what was expected in the initial elicitation, the revised priors were only modestly different from the initial priors. However, dramatic improvements were seen in coherency. For the five (out of these seven) experts who initially had RMSD values larger than 6m, four experts achieved a revised RMSD value of less than 4m.

\begin{table}[H]
\begin{threeparttable}[b]
  \centering
    \caption{Comparison of Initial and Revised means, SDs, and coherencies (RMSDs) of the inferred normal prior distributions for each expert (in meters)}
%  \resizebox{\textwidth}{!}{
  \begin{tabular}{lrrrrrr}
    \hline
    & \multicolumn{2}{c}{\textbf{Inferred Mean}}
    & \multicolumn{2}{c}{\textbf{Inferred SD}}
    & \multicolumn{2}{c}{\textbf{RMSD}} \\
    \cline{2-7}
    & \textbf{Initial} & \textbf{Revised} & \textbf{Initial} & \textbf{Revised} & \textbf{Initial} & \textbf{Revised} \\ \hline
    \textbf{Expert 1}  & $-0.3$  & 0.0   & 0.2   & 0.0    & 1.2 & 0.0 \\ \hline
    \textbf{Expert 2\tnote{*}} & $-0.3$  & $-0.3$  & 10.4  & 10.4 & 1.3 & 1.3 \\ \hline
    \textbf{Expert 3}  & $-30.8$ & $-30.9$ & 4.3   & 5.1  & 1.7 & 1.1 \\ \hline
    \textbf{Expert 4\tnote{*}} & $-9.4$  & $-9.4$  & 11.6  & 11.6 & 3.4 & 3.4 \\ \hline
    \textbf{Expert 5\tnote{*}} & 3.8   & 3.8   & 20.9  & 20.9 & 5.5 & 5.5 \\ \hline
    \textbf{Expert 6}  & 0.0     & 0.0     & 32.2  & 40.9 & 6.0 & 3.9 \\ \hline
    \textbf{Expert 7}  & 0.0     & 0.0     & 15.2  & 7.4  & 6.1 & 1.6 \\ \hline
    \textbf{Expert 8}  & $-2.6$  & $-3.3$  & 30.6  & 32.1 & 6.7 & 6.3 \\ \hline
    \textbf{Expert 9}  & $-15.8$ & $-12$   & 16.0    & 15.5 & 8.7 & 2.8 \\ \hline
    \textbf{Expert 10} & 0.9   & $-26.7$  & 118.6 & 21.3 & 21.7 & 3.7 \\ \hline
  \end{tabular}
  \begin{tablenotes}
       \item [*] Expert who did not submit revised responses
     \end{tablenotes}
      \end{threeparttable}
 % }
\end{table}

%\begin{figure}[H]
%    \centering
%    \includegraphics[width=1\linewidth]{b4andafter.pdf}
%    \caption{The elicited prior distribution before and %after the revision for each expert}
%\end{figure}

\section{Discussion}

We have described and pilot-tested a new prior elicitation approach that addresses two hurdles to obtaining credible prior distributions. First, we showed how our process for inferring the prior distribution from eliciting judgments about an expert's envisioned posterior quantities given hypothetical outcome data also provides an assessment of prior-posterior coherence, which can then be used by the experts to improve the coherency of their judgments. Second, through asking experts for judgments about expected values (i.e., point estimates), rather than probabilistic statements about uncertainty, we reduced their cognitive burden and avoided potential errors that might arise from misunderstanding of probability concepts. The case study results provide a \textit{proof-in-principle} demonstration of the feasibility of the proposed approach.

Although our results are promising, considerable additional development will be needed to make this approach broadly and easily accessible. First, the technical framework needs to be extended. In our case study, we considered only the elicitation of a single mean parameter from a normal data model and assumed that the expert's prior uncertainty could be captured using a conjugate normal distribution. This provided an approximate analytical expression for the posterior distribution and simplified the computations. However, for more complex problems involving multiple parameters or other types of outcomes (e.g., survival outcomes), conjugate priors will be unavailable and will require the use of more complex computational algorithms (e.g., Markov Chain Monte Carlo techniques, for example) to obtain posterior quantities \cite{hansen_practical_2018}. In addition, further exploration is needed into whether RMSD is the most appropriate metric for representing the degree of coherency and what constitutes adequate coherency. 

Second, elicitation is not purely a mathematical exercise but is a process in which the results are strongly influenced by how the process is implemented. At this time, the optimal elicitation process remains unclear on multiple facets. Small changes to the presentation of materials can impact the outcomes. As an example, during the revision stage, we changed the presentation of our scenarios from stand-alone questions only and added a table displaying the responses from all scenarios. Several experts commented that this helped them to see much more easily where their responses were not coherent, and assisted them when revising their judgments.  Other observations raise similar concerns. For example, how much guidance should be given to the experts before the initial elicitation? The poor coherence obtained in the initial elicitation suggests that the experts could have benefited from more guidance than was given in this study.  Yet, giving too much information initially could overwhelm the experts leading them to be uncomfortable with making any judgments. Nonetheless, the marked improvement achieved in the revised elicitation suggests that providing experts with a summary of common issues (like those identified in this study) might help them achieve a higher level of coherence from the outset. Obtaining reasonable responses at the outset is important to pre-empting anchoring bias, wherein experts are reluctant to revise initial responses even when they are shown to be incoherent.  Messaging to experts that their initial judgments should be treated as "tentative" might have pre-empted reports from some experts of feeling pressured to revise their initial responses, and clearer communication that the revision process is meant to provide an opportunity to reflect on whether their initial responses had captured their beliefs accurately might have made the task less stressful to the experts. More consultation with experts on how feedback should be presented and on the choice of language is needed to determine how to avoid such cognitive biases. Addressing these questions is pivotal for refining our methodology and improving the overall quality of the research outcomes. Future research may want to focus on these aspects and provide more structured elicitation methods for more robust and reliable results. Last, one challenge of elicitation is a lack of practical tools for fitting the appropriate mathematical models \cite{stefan_practical_2022,mikkola_prior_2023}.

In this study, we did not seek to compare the results of our new approach with what prior distributions would have been obtained if another existing elicitation approach had been used. While such comparisons would provide some indication of the sensitivity of elicited priors to the elicitation approach, judging whether one approach is ``better" than another is inherently difficult because no standard criteria for making such comparisons have been established. For example, one might consider assessing prior-posterior coherence for a prior distribution obtained by direct elicitation by eliciting posterior judgments as described here and then computing the coherency metric associated with this directly elicited prior. But, of course, this will always show poorer coherence than our approach since our procedure by construction selects the prior that maximizes coherency.

While we have illustrated our approach within the context of a stepped-wedge cluster randomized trial, the proposed method is applicable to any context that requires elicitation of a prior distribution, such as for other clinical trial designs or even in non-trial settings, such as health economics, where informed decision-making under uncertainty is essential. In particular, expert elicitation can be especially useful in early-phase trials, where direct evidence is limited and sample sizes are small \cite{robinson_using_2024,ROBINSON2025107918,bosma_proportional_2024}. For example, many targeted and immunotherapies are now applied across multiple cancer types due to shared molecular alterations.  In these settings, Bayesian elicitation is a particularly valuable tool, allowing the formal integration of expert opinion and historical evidence to guide dose selection, endpoint estimation, and trial design, ultimately improving efficiency and reducing uncertainty.

\section{Conclusions}
In conclusion, our PfP elicitation method has the potential to improve the trustworthiness of prior distributions by ensuring prior-posterior coherence and reducing the expert's cognitive burden during elicitation.  Further development is needed to expand the scope to inferring priors for parameter sets typically encountered in clinical trials.  These developments will help accelerate the adoption of Bayesian methods in clinical trials, leading to more robust and credible trial inferences.

\section*{List of abbreviations}
\textbf{PfP:} Prior from posterior \\
\textbf{RMSD:} Root mean square discrepancy\\
\textbf{WnW:} Walk 'n Watch\\
\textbf{SW-CRT:} Stepped-wedge cluster randomized trials\\
\textbf{MWT:} Minute walk test

\newpage
\section*{Declarations}

\subsection*{Ethics approval and consent to participate}
Not Applicable

\subsection*{Consent for publication}
All authors have reviewed and aggreed on the final version of the manuscript

\subsection*{Availability of data and materials}
No new data was generated for this study

\subsection*{Competing interests}
All authors declare no competing interests

\subsection*{Funding}
The CanStroke Recovery Trials Platform is supported by the Canada Brain Research Fund (CBRF), an innovative arrangement between the Government of Canada (through Health Canada) and Brain Canada Foundation and the Heart and Stroke Foundation Canadian Partnership for Stroke Recovery, University of Calgary, and the Sunnybrook Research Institute.

\subsection*{Authors' contributions}
HW and YO both conceived and developed the idea. HW contributed to the manuscript's revision, while YO led the manuscript writing. JE is the PI of the case study and facilitated the elicitation between statisticians (HW and YO) and experts (The WnW Research Team). JE and DZ provided significant contributions to the manuscript development. The WnW Research Team participated in the elicitation exercise. 

\subsection*{Acknowledgments}
The authors gratefully acknowledge the support and collaboration of the CanStroke Recovery Trials Platform.

\newpage
\bibliographystyle{ieeetr}
\bibliography{bib}\setlength{\itemsep}{0mm}

\newpage
\section*{Supplementary Materials}
\section*{S1: Walk n Watch secular trend elicitation questionnaire}

Name: \underline{\hspace{5in}}\\

The purpose of this exercise is to elicit your judgments about how much the efficacy of usual care changed during the Walk 'n Watch trial from year 2021 to year 2022. Many things could have changed the efficacy, including improvements in practice (positive) or staffing shortages (negative). You will be asked to state your estimate of the true change in efficacy in the study population (patients eligible for the Walk ‘n Watch trial) given various hypothetical outcomes in a sample of data. That is, we are interested in what you would conclude if the hypothetical outcomes presented here had been the actual data observed in the trial. (For this exercise, please disregard any actual trial data that you may have seen.)

There is no "right answer" -- rather, your response should reflect an assessment that combines both (a) your prior beliefs derived from past clinical data and experiences over 2021 and 2022, and (b) the sample data, according to how much weight you give to each component. For example, in one of the scenarios below, the sample data show an increase of 30 metres based on 10 patients in 2021 and 10 patients in 2022. But if you felt strongly a priori that the true change is near zero, say, and that this magnitude of change seen in the data is implausibly large (and perhaps primarily reflecting noise arising from the small sample), you should feel free to give little weight to the sample data and to choose a much smaller change or no change (zero) as your response. In contrast, in the scenario where the same 30 metres increase is based on 100 patients in 2021 and 100 patients in 2022, you may wish to give more weight to the data, and your response should be adjusted accordingly. We will ask you to consider 16 scenarios involving varying amounts of data and a variety of outcomes.

\subsection*{Scenarios}

For the first scenario, assume you have no sample data.

\begin{enumerate}[label=\textbf{\arabic*.}]
    \item By how much do you think the average improvement changed in the study population (that is, stroke patients eligible for the Walk ‘n Watch study) from 2021 to 2022?

    Response: \underline{\hspace{2in}} metres

    Please note, 0 metres would indicate that the clinical practice and improvements in patients was similar from 2021 to 2022; a positive value would indicate that practice had improved in 2022 resulting in an increase of the amount of metres in the 6MWT; a negative value would indicate that the practice had deteriorated in 2022 resulting in a reduction in the 6MWT.
    \end{enumerate}

\noindent
For the next 5 scenarios, the sample consists of 10 stroke patients from each year.

\begin{enumerate}[resume,label=\textbf{\arabic*.}]
    \item Suppose that in your random sample of 10 usual care stroke patients in 2021, their average improvement in 6MWT was 140 metres. Suppose also that in your random sample of 10 usual care stroke patients in 2022, their average improvement was 140 metres (no change compared to 2021).

    By how much do you think the average improvement changed in the study population (that is, stroke patients eligible for the Walk ‘n Watch study) from 2021 to 2022?

    Response: \underline{\hspace{2in}} metres

    Please note, 0 metres would indicate that the clinical practice and improvements in patients was similar from 2021 to 2022; a positive value would indicate that practice had improved in 2022 resulting in an increase of the amount of metres in the 6MWT; a negative value would indicate that the practice had deteriorated in 2022 resulting in a reduction in the 6MWT.
    
    \item Suppose that in your random sample of 10 usual care stroke patients in 2021, their average improvement in 6MWT was 140 metres. Suppose also that in your random sample of 10 usual care stroke patients in 2022, their average improvement was 150 metres (10 metre greater improvement compared to 2021).

    By how much do you think the average improvement changed in your population (that is, stroke patients eligible for the Walk ‘n Watch study) from 2021 to 2022?

    Response: \underline{\hspace{2in}} metres

    Please note, 0 metres would indicate that the clinical practice and improvements in patients was similar from 2021 to 2022; a positive value would indicate that practice had improved in 2022 resulting in an increase of the amount of metres in the 6MWT; a negative value would indicate that the practice had deteriorated in 2022 resulting in a reduction in the 6MWT.
    
    \item Suppose that in your random sample of 10 usual care stroke patients in 2021, their average improvement in 6MWT was 140 metres. Suppose also that in your random sample of 10 usual care stroke patients in 2022, their average improvement was 170 metres (30 metre greater improvement compared to 2021).

    By how much do you think the average improvement changed in your population (that is, stroke patients eligible for the Walk ‘n Watch study) from 2021 to 2022?

    Response: \underline{\hspace{2in}} metres

    Please note, 0 metres would indicate that the clinical practice and improvements in patients was similar from 2021 to 2022; a positive value would indicate that practice had improved in 2022 resulting in an increase of the amount of metres in the 6MWT; a negative value would indicate that the practice had deteriorated in 2022 resulting in a reduction in the 6MWT.
    
    \item Suppose that in your random sample of 10 usual care stroke patients in 2021, their average improvement in 6MWT was 140 metres. Suppose also that in your random sample of 10 usual care stroke patients in 2022, their average improvement was 130 metres (10 metre reduction compared to 2021).

    By how much do you think the average improvement changed in your population (that is, stroke patients eligible for the Walk ‘n Watch study) from 2021 to 2022?

    Response: \underline{\hspace{2in}} metres

    Please note, 0 metres would indicate that the clinical practice and improvements in patients was similar from 2021 to 2022; a positive value would indicate that practice had improved in 2022 resulting in an increase of the amount of metres in the 6MWT; a negative value would indicate that the practice had deteriorated in 2022 resulting in a reduction in the 6MWT.
    
    \item Suppose that in your random sample of 10 usual care stroke patients in 2021, their average improvement in 6MWT was 140 metres. Suppose also that in your random sample of 10 usual care stroke patients in 2022, their average improvement was 110 metres (30 metre reduction compared to 2021).

    By how much do you think the average improvement changed in your population (that is, stroke patients eligible for the Walk ‘n Watch study) from 2021 to 2022?

    Response: \underline{\hspace{2in}} metres

    Please note, 0 metres would indicate that the clinical practice and improvements in patients was similar from 2021 to 2022; a positive value would indicate that practice had improved in 2022 resulting in an increase of the amount of metres in the 6MWT; a negative value would indicate that the practice had deteriorated in 2022 resulting in a reduction in the 6MWT.
        \end{enumerate}

\noindent
For the next 5 scenarios, the sample consists of 30 stroke patients from each year.

\begin{enumerate}[resume,label=\textbf{\arabic*.}]
    
    \item Suppose that in your random sample of 30 usual care stroke patients in 2021, their average improvement in 6MWT was 140 metres. Suppose also that in your random sample of 30 usual care stroke patients in 2022, their average improvement was 140 metres (no change compared to 2021).

    By how much do you think the average improvement changed in the study population (that is, stroke patients eligible for the Walk ‘n Watch study) from 2021 to 2022?

    Response: \underline{\hspace{2in}} metres

    Please note, 0 metres would indicate that the clinical practice and improvements in patients was similar from 2021 to 2022; a positive value would indicate that practice had improved in 2022 resulting in an increase of the amount of metres in the 6MWT; a negative value would indicate that the practice had deteriorated in 2022 resulting in a reduction in the 6MWT.
    
    \item Suppose that in your random sample of 30 usual care stroke patients in 2021, their average improvement in 6MWT was 140 metres. Suppose also that in your random sample of 30 usual care stroke patients in 2022, their average improvement was 150 metres (10 metre greater improvement compared to 2021).

    By how much do you think the average improvement changed in your population (that is, stroke patients eligible for the Walk ‘n Watch study) from 2021 to 2022?

    Response: \underline{\hspace{2in}} metres

    Please note, 0 metres would indicate that the clinical practice and improvements in patients was similar from 2021 to 2022; a positive value would indicate that practice had improved in 2022 resulting in an increase of the amount of metres in the 6MWT; a negative value would indicate that the practice had deteriorated in 2022 resulting in a reduction in the 6MWT.
    
    \item Suppose that in your random sample of 30 usual care stroke patients in 2021, their average improvement in 6MWT was 140 metres. Suppose also that in your random sample of 30 usual care stroke patients in 2022, their average improvement was 170 metres (30 metre greater improvement compared to 2021).

    By how much do you think the average improvement changed in your population (that is, stroke patients eligible for the Walk ‘n Watch study) from 2021 to 2022?

    Response: \underline{\hspace{2in}} metres

    Please note, 0 metres would indicate that the clinical practice and improvements in patients was similar from 2021 to 2022; a positive value would indicate that practice had improved in 2022 resulting in an increase of the amount of metres in the 6MWT; a negative value would indicate that the practice had deteriorated in 2022 resulting in a reduction in the 6MWT.
    
    \item Suppose that in your random sample of 30 usual care stroke patients in 2021, their average improvement in 6MWT was 140 metres. Suppose also that in your random sample of 30 usual care stroke patients in 2022, their average improvement was 130 metres (10 metre reduction compared to 2021).

    By how much do you think the average improvement changed in your population (that is, stroke patients eligible for the Walk ‘n Watch study) from 2021 to 2022?

    Response: \underline{\hspace{2in}} metres

    Please note, 0 metres would indicate that the clinical practice and improvements in patients was similar from 2021 to 2022; a positive value would indicate that practice had improved in 2022 resulting in an increase of the amount of metres in the 6MWT; a negative value would indicate that the practice had deteriorated in 2022 resulting in a reduction in the 6MWT.
    
    \item Suppose that in your random sample of 30 usual care stroke patients in 2021, their average improvement in 6MWT was 140 metres. Suppose also that in your random sample of 30 usual care stroke patients in 2022, their average improvement was 110 metres (30 metre reduction compared to 2021).

    By how much do you think the average improvement changed in your population (that is, stroke patients eligible for the Walk ‘n Watch study) from 2021 to 2022?

    Response: \underline{\hspace{2in}} metres

    Please note, 0 metres would indicate that the clinical practice and improvements in patients was similar from 2021 to 2022; a positive value would indicate that practice had improved in 2022 resulting in an increase of the amount of metres in the 6MWT; a negative value would indicate that the practice had deteriorated in 2022 resulting in a reduction in the 6MWT.
    \end{enumerate}

\noindent
For the next 5 questions, the sample consists of 100 stroke patients from each year.

\begin{enumerate}[resume, label=\textbf{\arabic*.}]
    \item Suppose that in your random sample of 100 usual care stroke patients in 2021, their average improvement in 6MWT was 140 metres. Suppose also that in your random sample of 100 usual care stroke patients in 2022, their average improvement was 140 metres (no change compared to 2021).

    By how much do you think the average improvement changed in the study population (that is, stroke patients eligible for the Walk ‘n Watch study) from 2021 to 2022?

    Response: \underline{\hspace{2in}} metres

    Please note, 0 metres would indicate that the clinical practice and improvements in patients was similar from 2021 to 2022; a positive value would indicate that practice had improved in 2022 resulting in an increase of the amount of metres in the 6MWT; a negative value would indicate that the practice had deteriorated in 2022 resulting in a reduction in the 6MWT.
    
    \item Suppose that in your random sample of 100 usual care stroke patients in 2021, their average improvement in 6MWT was 140 metres. Suppose also that in your random sample of 100 usual care stroke patients in 2022, their average improvement was 150 metres (10 metre greater improvement compared to 2021).

    By how much do you think the average improvement changed in your population (that is, stroke patients eligible for the Walk ‘n Watch study) from 2021 to 2022?

    Response: \underline{\hspace{2in}} metres

    Please note, 0 metres would indicate that the clinical practice and improvements in patients was similar from 2021 to 2022; a positive value would indicate that practice had improved in 2022 resulting in an increase of the amount of metres in the 6MWT; a negative value would indicate that the practice had deteriorated in 2022 resulting in a reduction in the 6MWT.
    
    \item Suppose that in your random sample of 100 usual care stroke patients in 2021, their average improvement in 6MWT was 140 metres. Suppose also that in your random sample of 100 usual care stroke patients in 2022, their average improvement was 170 metres (30 metre greater improvement compared to 2021).

    By how much do you think the average improvement changed in your population (that is, stroke patients eligible for the Walk ‘n Watch study) from 2021 to 2022?

    Response: \underline{\hspace{2in}} metres

    Please note, 0 metres would indicate that the clinical practice and improvements in patients was similar from 2021 to 2022; a positive value would indicate that practice had improved in 2022 resulting in an increase of the amount of metres in the 6MWT; a negative value would indicate that the practice had deteriorated in 2022 resulting in a reduction in the 6MWT.
    
    \item Suppose that in your random sample of 100 usual care stroke patients in 2021, their average improvement in 6MWT was 140 metres. Suppose also that in your random sample of 100 usual care stroke patients in 2022, their average improvement was 130 metres (10 metre reduction compared to 2021).

    By how much do you think the average improvement changed in your population (that is, stroke patients eligible for the Walk ‘n Watch study) from 2021 to 2022?

    Response: \underline{\hspace{2in}} metres

    Please note, 0 metres would indicate that the clinical practice and improvements in patients was similar from 2021 to 2022; a positive value would indicate that practice had improved in 2022 resulting in an increase of the amount of metres in the 6MWT; a negative value would indicate that the practice had deteriorated in 2022 resulting in a reduction in the 6MWT.
    
    \item Suppose that in your random sample of 100 usual care stroke patients in 2021, their average improvement in 6MWT was 140 metres. Suppose also that in your random sample of 100 usual care stroke patients in 2022, their average improvement was 110 metres (30 metre reduction compared to 2021).

    By how much do you think the average improvement changed in your population (that is, stroke patients eligible for the Walk ‘n Watch study) from 2021 to 2022?

    Response: \underline{\hspace{2in}} metres

    Please note, 0 metres would indicate that the clinical practice and improvements in patients was similar from 2021 to 2022; a positive value would indicate that practice had improved in 2022 resulting in an increase of the amount of metres in the 6MWT; a negative value would indicate that the practice had deteriorated in 2022 resulting in a reduction in the 6MWT.
\end{enumerate}

\newpage

\section*{S2: Example of individual feedback}
Here is an example of individual feedback for expert 9

\begin{figure}[H]
    \centering
    \includegraphics[width=1\linewidth]{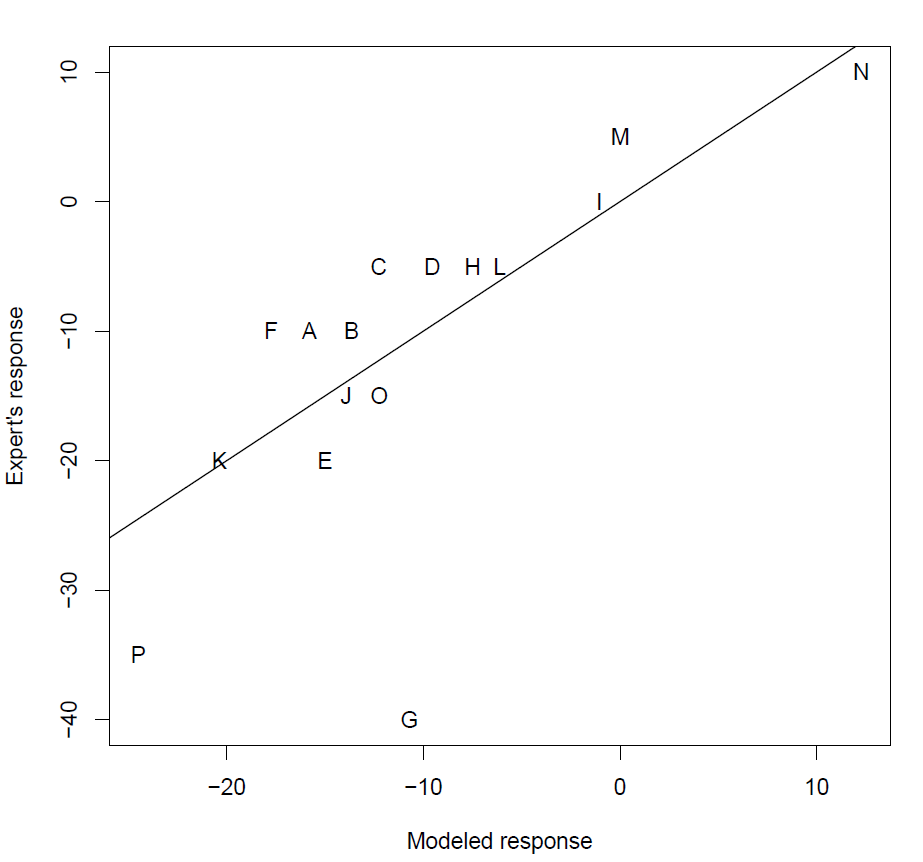}
    %\caption{}
\end{figure}

\subsection*{Individual feedback}
The fitted model indicates that you expect the one-year change to be a decrease of 15.8 meters with a standard deviation (SD) of 16 meters.  Your responses largely are consistent with each other and with the predicted response from the model, as shown in the plot, though there are some scenarios which do not fit well.  If you had been perfectly consistent, then your responses and the modeled response would fall on the straight line shown. The SD of 16 reflects the estimate of your (un)certainty about the one-year change.  Roughly, it means that you are ~95\% sure that the true one-year change is in the range (-48, 16).  (Note that range will change (likely to become narrower) if you decide to revise your responses to achieve greater consistency.)

The exact values of the points on the plot are shown in the table below.  The rows highlighted in red are the scenarios in which your response (column ‘Expert’) is highly inconsistent because it falls outside the range between your initial judgment prior to seeing any data (cell in green) and the observed result in the data (column ‘Observed mean change’), and it also is far from what the value predicted by the model. We ask that you reflect on and revise your responses for these scenarios and/or your initial judgment (if you feel it no longer represents your beliefs) to resolve these discrepancies as they have a major impact on the results. The rows highlighted in yellow are scenarios that are modestly inconsistent, so revising them would improve consistency but would have less impact on the results. Examples:

In Scenario 7 (`G’), your response was -40.  However, the data suggest no change, so if you keep your initial judgment of -10, a consistent response would be between 0 and -10.

In Scenario 5 (`E’), your response was -20.  The data suggest a change of -10, which exactly supports your initial judgment, so it would make sense to choose -10 as your response.  However, because your response is still somewhat close to the consistent choice, revising to -10 likely won’t have a great impact on the final result.

\begin{table}[H]
\centering
\resizebox{\textwidth}{!}{
\begin{tabular}{|c|c|c|c|c|c|c|}
\hline
\textbf{Scenario} & \textbf{Plot Symbol} & \textbf{n} & \textbf{Observed Mean Change} & \textbf{Expert Response} & \textbf{Model-fitted Responses} & \textbf{Difference}  \\ \hline

1  & A  & NA  & NA  & -10   & -15.8  & 5.8  \\ \hline
2  & B  & 10  & 0   & -10   & -13.6  & 3.6  \\ \hline
3  & C  & 10  & +10 & -5    & -12.3  & 7.3  \\ \hline
4  & D  & 10  & +30 & -5    & -9.5   & 4.5  \\ \hline
5  & E  & 10  & -10 & -20   & -15    & -5   \\ \hline
6  & F  & 10  & -30 & -10   & -17.7  & 7.7  \\ \hline
7  & G  & 30  & 0   & -40   & -10.7  & -29.3 \\ \hline
8  & H  & 30  & +10 & -5    & -7.5   & 2.5  \\ \hline
9  & I  & 30  & +30 & 0     & -1.1   & 1.1  \\ \hline
10 & J  & 30  & -10 & -15   & -13.9  & -1.1 \\ \hline
11 & K  & 30  & -30 & -20   & -20.4  & 0.4  \\ \hline
12 & L  & 100 & 0   & -5    & -6.1   & 1.1  \\ \hline
13 & M  & 100 & +10 & 5     & 0      & 5    \\ \hline
14 & N  & 100 & +30 & 10    & 12.3   & -2.3 \\ \hline
15 & O  & 100 & -10 & -15   & -12.2  & -2.8 \\ \hline
16 & P  & 100 & -30 & -35   & -24.5  & -10.5 \\ \hline
\end{tabular}
}
\end{table}

\end{abstract}\end{document}